\documentclass[journal,10pt]{IEEEtran}

\usepackage[T1]{fontenc}
\usepackage{amsfonts}
\usepackage{epsf}
\usepackage{graphicx}
\usepackage{algorithm}
\usepackage{algpseudocode}
\usepackage{subfigure}
\usepackage{amssymb,amsmath,amsthm}
\usepackage{cite}
\usepackage{color}

\newtheorem{theorem}{Theorem}

\newcommand{\norm}[1]{\left\lVert#1\right\rVert}

\allowdisplaybreaks \allowdisplaybreaks[4]
\begin{document}
\title{\LARGE Trajectory Design for Distributed Estimation in UAV Enabled Wireless Sensor Network}
\author{Cheng~Zhan,~\IEEEmembership{Member,~IEEE,}
        Yong~Zeng,~\IEEEmembership{Member,~IEEE,}
        and~Rui~Zhang,~\IEEEmembership{Fellow,~IEEE}

\thanks{C. Zhan is with the School of Computer and Information Science, Southwest University,
Chongqing 400715, China. (e-mail: zhanc@swu.edu.cn).}
\thanks{Y. Zeng and R. Zhang are with the Department of Electrical and Computer
Engineering, National University of Singapore, Singapore 117583 (e-mail:
\{elezeng, elezhang\}@nus.edu.sg).}}

\maketitle
\thispagestyle{empty}

\begin{abstract}
In this paper, we study an unmanned aerial vehicle (UAV)-enabled wireless sensor network, where a UAV is dispatched to collect the sensed data from distributed sensor nodes (SNs) for estimating an unknown parameter. It is revealed that in order to minimize the mean square error (MSE) for the estimation, the UAV should collect the data from as many SNs as possible, based on which an optimization problem is formulated to design the UAV's trajectory subject to its practical mobility constraints. Although the problem is non-convex and NP-hard, we show that the optimal UAV trajectory consists of connected line segments only. With this simplification, an efficient suboptimal solution is proposed by leveraging the classic traveling salesman problem (TSP) method and applying convex optimization techniques. Simulation results show that the proposed trajectory design achieves significant performance gains in terms of the number of SNs whose data are successfully collected, as compared to other benchmark schemes.
\end{abstract}


\section{Introduction}\label{intro}
The last two decades have witnessed a dramatic advancement in the
research and development of wireless sensor network (WSN)
for applications in various fields. A WSN typically consists of a large number of sensor nodes (SNs) that are distributed in a wide area of interest. SNs are typically low-cost and low-power devices, which are able to sense, process,
store and transmit information. Although the SNs usually have limited sensing, processing and transmission capabilities individually, their collaborative estimation/detection can be highly efficient and reliable \cite{JXiaoLuo,JXiao}.

One typical application of WSN is for the estimation of an unknown parameter (such as pressure, temperature, etc.) in a given field based on noisy observations collected from distributed SNs. Specifically, each
SN performs local sensing and signal quantization, then sends the quantized data to a Fusion Center (FC),
where the received data from all SNs are jointly processed to produce a final estimate of the unknown parameter. Prior research on distributed estimation in WSN (see, e.g., \cite{JXiaoLuo,JXiao}) has mainly considered the static FC at a fixed location. As a result, SNs may require significantly different transmission power to send their data reliably to the FC  due to their near-far distances from it, which results in inhomogeneous energy consumption rates of the SNs and thus limited lifetime of the WSN.

To overcome this issue, utilizing unmanned aerial vehicle
(UAV) as a mobile data collector for WSN has been proposed as a promising solution \cite{YZengmag,SSay,CZhan}. With on-board miniaturized transceivers that enable ground-to-air communications, UAV-enabled
WSN has promising
advantages, such as the ease of on-demand and swift deployment,
the flexibility with fully-controllable mobility, and the high
probability of having line-of-sight (LoS) communication links
with the ground SNs. In contrast to fixed FCs, a UAV-enabled mobile data collector is able to fly sufficiently close to each SN to collect its sensed data more reliably, thus helping significantly reduce the SNs' energy consumptions, yet in a more fair manner.\vspace{-0.2in}
\begin{figure}[ht]\centering
\includegraphics[width=2.6in]{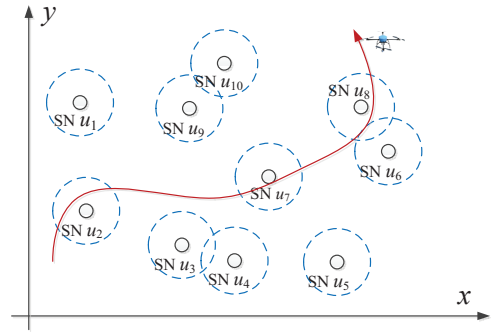}
\caption{A UAV-enabled mobile data collector for wireless sensor network.}
\label{example}\vspace{-0.1in}
\end{figure}

A fundamental problem
in UAV-enabled WSN for distributed estimation is the design of the UAV's trajectory (see Fig. \ref{example}), which needs to take into account two important considerations. Firstly, for an SN to send its data reliably to the UAV, the UAV needs to fly sufficiently close to the SN (say, within a certain maximum distance assuming an LoS channel between them). Secondly, given a finite flight duration,
the UAV's trajectory should be designed to ``cover'' (with respect to the given maximum distance) as many SNs as possible to optimize the distributed estimation performance (e.g., minimizing the mean square error (MSE) for the estimated parameter). Notice that in our prior work \cite{CZhan}, the SNs' wakeup schedule and UAV's trajectory were jointly optimized to minimize the maximum energy consumption of all SNs, while ensuring that the required amount of data is collected reliably from each SN. In contrast to \cite{CZhan} where the UAV needs to collect independent data from all SNs, this paper considers that all SNs' data contains noisy observations of a common unknown parameter. As a result, their approaches for the UAV trajectory design are also fundamentally different.

UAV trajectory design for optimizing communication performance has received growing interests recently (see. e.g., \cite{YZengRelay,QWu,YZengEner,YZeng2,MMozaffari}). In \cite{YZengRelay}, the UAV's trajectory was jointly optimized with transmission power/rate for throughput
maximization in a UAV-enabled mobile relaying system, subject to practical mobility constraints of the UAV. The energy-efficient
UAV communication via optimizing the
UAV's trajectory was studied in \cite{YZengEner}, which aims to strike an optimal balance between maximizing the communication rate and minimizing the UAV's propulsion power
consumption. The deployment and movement of multiple UAVs, used as aerial base stations to collect data from ground Internet of Things (IoT) devices, was investigated in \cite{MMozaffari}. The work in \cite{QWu} maximized the minimum throughput of a multi-UAV-enabled wireless network by optimizing the multiuser communication scheduling jointly with the UAVs' trajectory
and power control. In \cite{YZeng2}, the UAV trajectory was designed to minimize the mission completion time for UAV-enabled multicasting.  Different from the above work, this paper investigates the UAV trajectory design under a new setup for distributed estimation in WSN. The main contributions of this paper are summarized as follows:
\begin{itemize}
\item
First, we show that for distributed estimation in an UAV-enabled WSN, minimizing the MSE is equivalent to maximizing the number of SNs whose sensed data are reliably collected by the UAV;
\item
Second, with a given UAV flight duration, we formulate an optimization problem for designing the UAV's trajectory to maximize the number of covered  SNs, subject to the practical constraints on the initial and final locations of the UAV as well as its maximum speed. Although the problem is NP-hard, we show that the optimal UAV trajectory consists of connected line segments only;
\item
Third, with the above simplification, an efficient greedy algorithm is proposed to obtain a high-quality suboptimal trajectory solution by leveraging the classic traveling salesman problem (TSP) method and applying convex optimization techniques;
\item
Last, numerical results show that the proposed trajectory design achieves significant performance gains in terms of the number of SNs with successful data collection as compared to benchmark schemes.
\end{itemize}\vspace{-0.1in}

\section{System Model and Problem Formulation}\label{model}
As shown in Fig. \ref{example}, we consider a WSN consisting of $N$ SNs arbitrarily located on the ground, denoted by $U=\{u_1,u_2,\ldots,u_N\}$. The horizontal coordinate of SN $u_n$ is
denoted by $\mathbf{w}_n\in \mathbb{R}^{2\times 1}$, $n=1, \cdots, N$. Each SN can observe,
quantize and transmit its observation for an unknown parameter $\theta$ to the FC, which estimates the
parameter based on the received information.\vspace{-0.15in}
\subsection{Distributed Estimation}
Each SN $u_i$ makes a noisy observation on a deterministic parameter $\theta$ (e.g., temperature). The real-value observation $y_i$ by SN $u_i$ is modeled as
\begin{eqnarray}
y_i =\theta+n_i,
\end{eqnarray}
where $n_i$ is the observation noise that is assumed to be spatially uncorrelated for different SNs with zero mean and variance $\sigma_i^2$. We further assume that the noise variances for all SNs are identical, i.e., $\sigma_i^2=\sigma^2, \forall i$. Denote by $[-W,W]$ the signal range that the sensors can observe, where $W$ is a known constant that is typically determined by the sensor's dynamic range. In other words, $y_i\in [-W,W]$.

The local processing at SN $u_i$ consists of the following: (i) an
uniform quantizer with $2^{S_i}$ quantization levels, where $S_i$ denotes the number of quantization bits and $\Delta_i=\frac{2W}{2^{S_i}-1}$ represents the
quantization step size; (ii) a modulator, which maps the $S_i$ quantization bits into a number of symbols based on certain modulation scheme, such as binary phase shift keying (BPSK); and (iii) transmission of the modulated symbols to the FC. It is shown in \cite{ASani} that with uniform quantizer, the quantization noise variance for $u_i$ can be obtained as $\delta_i^2=\frac{\Delta_i^2}{12}$. For a homogeneous sensor network with equal observation noise power for all SNs, we assume that all SNs generate the same number of quantization bits, i.e., $S_i=S, \forall i$ \cite{JXiaoLuo}. The FC then performs
the linear estimation based on the received data from all SNs to recover $\theta$ using the Quasi Best Linear Unbiased Estimators (Quasi-BLUE) \cite{JXiao}, and the corresponding MSE can be obtained as
\begin{eqnarray}
\text{MSE}=\left(\sum_{i=1}^K\frac{1}{\sigma_i^2+\delta_i^2}\right)^{-1}=\frac{1}{K}\left(\sigma^2+\frac{W^2}{3(2^S-1))}\right),
\end{eqnarray}
where $K\leq N$ is the number of SNs whose sensed data are reliably collected.

The expression in (2) shows that for the considered distributed estimation, the MSE is inversely proportional to the number of SNs $K$ whose data are reliably collected. Therefore, in order to minimize MSE for the distributed estimation, the FC should successfully collect data from as many SNs as possible.\vspace{-0.15in}
\subsection{UAV Data Collection}
For the UAV-enabled WSN, a UAV is employed as a flying data collector/FC for a given time horizon $T$, which collects the quantized information from SNs and jointly estimates the parameter $\theta$. It is assumed that the UAV flies at a fixed altitude of $H$ in meter (m) and the maximum speed is denoted as $V_{\max}$ in meter/second (m/s). The initial and
final UAV horizontal locations are pre-determined and denoted as $\mathbf{q}_0, \mathbf{q}_F \in \mathbb{R}^{2\times 1}$, respectively, where $\norm{\mathbf{q}_F-\mathbf{q}_0}\leq V_{\max}T$ so that there exists at least one feasible trajectory for the UAV to fly from $\mathbf{q}_0$ to $\mathbf{q}_F$ in a straight line within $T$. The UAV's flying trajectory projected on the ground is denoted as $\mathbf{q}(t)\in \mathbb{R}^{2\times 1}, 0 \leq t \leq T$.

We assume that the transmit power for each SN is given (but can be different among SNs in general, depending on each SN's energy availability). Thus, in order to satisfy the
minimum required signal-to-noise ratio (SNR) at the UAV
for reliable data collection from each SN $u_n$, the UAV location projected on the ground should lie within its communication range, which is denoted by $r_n$. For each SN $u_n$, define the \textit{coverage area} $D_n\triangleq \{\mathbf{q}\in \mathbb{R}^{2\times1} \mid \norm{\mathbf{q}-\mathbf{w}_n}\leq r_n\}$. In general, an SN with smaller transmit power has a smaller $r_n$ given the same $S$ for all SNs. As a result, the UAV can collect the data reliably from $u_n$ as long as it is within $D_n$, as shown in Fig. \ref{example}. In the following, we refer to the event that the UAV enters into $D_n$ as {\em UAV visits} $u_n$. For example, in Fig. \ref{example}, the UAV has visited SNs $u_2$, $u_6$, $u_7$ and $u_8$. Since the number of quantization bits is typically small for practical applications (e.g., $S=10$ bits) \cite{JXiaoLuo}, the required transmission time for the quantized information can be neglected compared to the UAV flight time. In other words, as long as the UAV visits $u_n$, we assume that the sensed data by $u_n$ can be reliably collected by the UAV.\vspace{-0.1in}
\subsection{Problem Formulation}
Define the indicator function $\hat{I}_n(t)$ and indicator variable $I_n$ as follows,\vspace{-0.1in}
\begin{eqnarray}
\hat{I}_n(t)=
\begin{cases}
1,& \text{if } \mathbf{q}(t)\in D_n\\
0,& \text{otherwise}
\end{cases}\\
I_n=
\begin{cases}
1,& \text{if } \int_{0}^{T}{\hat{I}_n(t)dt}>0\\
0,& \text{otherwise}
\end{cases}
\end{eqnarray}
where $\hat{I}_n(t)$ indicates whether UAV is within $D_n$ or not at each time instant $t$, and ${I}_n$ indicates whether UAV visits $D_n$ (at least once) during the time horizon $T$. We assume that all the SNs' locations as well as their communication ranges are known to the UAV. The UAV trajectory design problem to maximize the number of visited SNs for distributed estimation is thus formulated as,\vspace{-0.1in}
\begin{eqnarray}
\text{(P1):} &&\max\limits_{\mathbf{q}(t)} \sum_{n=1}^{N}I_n\nonumber\\
\text{s.t.} &&\norm{\dot{\mathbf{q}}(t)}\leq V_{\max}, 0\leq t\leq T,\\
&&\mathbf{q}(0)=\mathbf{q}_0, \mathbf{q}(T)=\mathbf{q}_F.
\end{eqnarray}
In (P1), constraint (5) corresponds to the maximum UAV speed constraint, with $\dot{\mathbf{q}}(t)$ denoting the time-derivative of ${\mathbf{q}}(t)$, and constraints (6) specify the initial/final locations for the UAV.

\section{Proposed Solution}
(P1) is a non-convex optimization problem,
since the objective function is a non-concave function, which involves time-dependent
indicator functions in terms of the UAV trajectory. In the following, we first show the structure of the optimal UAV trajectory solution to (P1).\vspace{-0.15in}
\subsection{Optimal Trajectory Structure and Problem Reformulation}\label{structure}
\begin{theorem}
Without loss of optimality to problem (P1), the UAV trajectory can be assumed to consist of
connected line segments only.
 \label{useful}
\end{theorem}\vspace{0.05in}
\begin{proof}
Theorem \ref{useful} is proved by showing that for any given feasible
trajectory $\mathbf{q}(t)$ of (P1), which contains curved path, we can always construct another feasible trajectory $\mathbf{q}'(t)$ consisting of only
connected line segments, which satisfies the conditions
in (5), (6), and achieves the same objective value. Specifically, for any given $\mathbf{q}(t)$, the indicator variables $I_n$ can be obtained based on (3) and (4), and the objective value of (P1) (i.e., the number of visited SNs) can be obtained as $K=\sum_{n=1}^{N}I_{n}$. Without loss of generality, we assume that the $K$ visited SNs are $u_{\omega_1},u_{\omega_2},\ldots,u_{\omega_K}$, where ${\omega_i}$ is the index of the visited SNs in $U$, $1\leq i\leq K$. Let $\mathbf{q}_{\omega_i}$ be the waypoint that the UAV enters into $D_{\omega_i}$ for the first time, $1\leq i\leq K$, then $\mathbf{q}_{\omega_i}=\mathbf{q}(t_{\omega_i})$, where $t_{\omega_i}=\min\{t\mid \hat{I}_{\omega_i}(t)=1, 0\leq t\leq T\}$. We re-arrange $\mathbf{q}_{\omega_i}$ with the increasing order of $t_{\omega_i}$ and obtain a sequence of ordered waypoins $(\mathbf{q}_{\pi_1},\ldots,\mathbf{q}_{\pi_{K}})$, where $(\pi_1,\ldots,\pi_K)$ is a permutation of $(\omega_1,\ldots,\omega_K)$. Let $\mathbf{q}_{\pi_0}=\mathbf{q}_{0}$ and $\mathbf{q}_{\pi_{K+1}}=\mathbf{q}_{F}$. Then we have $T=\sum_{i=0}^{K}{T_{\pi_{i}\pi_{i+1}}}$, where $T_{\pi_{i}\pi_{i+1}}$ denotes the flying time between waypoints $\mathbf{q}_{\pi_i}$ and $\mathbf{q}_{\pi_{i+1}}$ along the given trajectory $\mathbf q(t)$. We can then replace any curved trajectory path between waypoints $\mathbf{q}_{\pi_i}$ and $\mathbf{q}_{\pi_{i+1}}$, $1\leq i\leq K$ with a line segment and obtain the alternative trajectory $\mathbf{q}'(t)$. Thus, with the same flying time $T_{\pi_{i}\pi_{i+1}}$, the required flying speed $\norm{\dot{\mathbf{q}'}(t)}$ can be reduced since line segment between any given pair of waypoints $\mathbf{q}_{\pi_i}$ and $\mathbf{q}_{\pi_{i+1}}$ yields the minimum distance. Therefore, $\mathbf{q}'(t)$ satisfies the constraints (5) and (6), and yet achieves the same objective value for (P1). This concludes the proof of Theorem 1.
\end{proof}

Based on Theorem 1 and its proof, (P1) can be solved by determining the optimal subset of SNs that are visited, denoted as $U_K\subseteq U$ with cardinality $K$, their optimal visiting order $\boldsymbol\pi=(\pi_1,\ldots,\pi_K)$, and the optimal waypoints $\mathbf{q}_{\pi_k}\in \mathbb{R}^{2\times 1}$ each for an SN $u_{\pi_k}\in U_K$, such that the data from $u_{\pi_k}$ can be received when the UAV is at $\mathbf{q}_{\pi_k}$ and the total distance of the resulting path $\mathbf{p}=(\mathbf{q}_{\pi_0},\dots,\mathbf{q}_{\pi_{K+1}})$ is no greater than $V_{\max}T$. Therefore, (P1) can be reformulated as
\begin{eqnarray}
\text{(P2):} \max\limits_{K,U_K,\mathbf{q}_{\pi_k},\boldsymbol\pi} && K\nonumber\\
\text{s.t.} && U_K\subseteq U, \\
&& \sum_{k=1}^{K+1}\norm{\mathbf{q}_{\pi_k}-\mathbf{q}_{\pi_{k-1}}}\leq V_{\max}T,\\
&& \norm{\mathbf{q}_{\pi_k}-\mathbf{w}_{\pi_k}}\leq r_{\pi_k}, 1\leq k\leq K.
\end{eqnarray}

Consider a special case of (P2) with $r_n=0, \forall n$, (i.e., the UAV can collect data reliably from the SN only when it is directly above the SN), then only $U_K$ and the visiting order $\boldsymbol\pi$ need to be determined to maximize the number of visited SNs within duration $T$. This problem is essentially equivalent to the selective TSP problem (or orienteering problem), which is known to be NP-hard \cite{PVansteenwegen}. Therefore, problem (P2) with $r_n\geq 0$ is also NP-hard and more difficult to solve than TSP in general.\vspace{-0.15in}
\subsection{Proposed Algorithm for (P2)}
One straightforward approach for solving (P2) is via exhaustively searching all possible subsets $U_K\subseteq U$ and the visiting order $\boldsymbol\pi$ of each $U_K$, and then determining whether the minimum distance of path $\mathbf{p}$ that visits $U_K$ with order $\boldsymbol\pi$ is no greater than $V_{\max}T$. However, searching all possible subsets of $U$ has an exponential complexity of $O(2^N)$, which is infeasible for large values of $N$. Therefore, we propose an efficient suboptimal solution to (P2) by a greedy iterative algorithm.

The key idea of our proposed solution is to maintain a working set $\mathcal{C}$ containing SNs that the UAV
needs to visit, and add only one additional SN in $\mathcal{C}$ at each iteration. Initially, $\mathcal{C}$ is set as empty, and we make a greedy choice to select $u_k$ from the complement set $U\setminus \mathcal{C}$, which leads to the minimum traveling distance to visit $\mathcal{C}\bigcup\{u_k\}$. The above process iterates until $\mathcal{C}=U$ or when the required visiting time is greater than $T$. Let $\mathbf{w}_0=\mathbf{q}_{0}$ and $\mathbf{w}_{N+1}=\mathbf{q}_{F}$. The proposed greedy algorithm for (P2) is summarized in Algorithm 1. In Algorithm 1, $d_{\max}$ is the flying distance of the UAV that flies over
each SN following the increasing index of all the SNs in $U$, which is an upper bound of the minimum flying distance to visit all SNs in $U$.

\begin{algorithm}[h]
\caption{Proposed algorithm for (P2)}
\begin{algorithmic}[1]
\State Initialize $\mathcal{C}\leftarrow \emptyset$;
\State $d_{\min}\leftarrow 0$; $d_{\max}\leftarrow \sum_{i=1}^{N+1}\norm{\mathbf{w}_{i}-\mathbf{w}_{i-1}}$;
\While {$\mathcal{C}\neq U$ and $d_{\min}<d_{\max}$}
\State $d_{\min}\leftarrow d_{\max}$; $U_K^{*}\leftarrow \mathcal{C}$;
\For {each $u_k\in U\setminus \mathcal{C}$};
\State $U_K\leftarrow \mathcal{C}\bigcup\{u_k\}$;
\State Given $U_K$, solve (P3) and denote the optimized \indent\quad objective value and trajectory as $d_0$ and $\mathbf{Q}_0$;
\If {$d_0\leq V_{\max}T$ and $d_0<d_{\min}$}
\State $d_{\min}\leftarrow d_0$; $\mathbf{Q}^{*}\leftarrow\mathbf{Q}_0$; $U_K^{*}\leftarrow U_K$;
\EndIf
\EndFor
\State $\mathcal{C}\leftarrow U_K^{*}$;
\EndWhile
\State\textbf{Output}:$\mathbf{Q}^{*}, U_K^*$;
\label{code:recentEnd}
\end{algorithmic}
\end{algorithm}

Note that in step 7 of Algorithm 1, the UAV trajectory is designed with a given SN set $U_K$ to minimize the UAV traveling distance. The problem is formulated as
\begin{eqnarray}
\text{(P3):} \min\limits_{\mathbf{q}_{\pi_k},\boldsymbol\pi} && \sum_{k=1}^{K+1}\norm{\mathbf{q}_{\pi_k}-\mathbf{q}_{\pi_{k-1}}}\nonumber\\
\text{s.t.} && \norm{\mathbf{q}_{\pi_k}-\mathbf{w}_{\pi_k}}\leq r_{\pi_k}, 1\leq k\leq K.
\end{eqnarray}

In Algorithm 1, after executing the inner iteration from step 5 to step 11, if adding any additional SN in the complementary
set $U\setminus \mathcal{C}$ leads to a traveling distance greater than $V_{\max}T$, then step 9 will not be executed and $d_{\min}$ remains equal to $d_{\max}$ as initialized in step 4, and the outer iteration in step 3 terminates. Otherwise, step 9 will be executed and one additional SN will be added into $\mathcal{C}$ with $d_{\min}<d_{\max}$, and the outer iteration continues. Therefore, the size of $\mathcal{C}$ increases over the iterations until either $\mathcal{C}=U$ or adding any additional SN will lead to a traveling distance greater than $V_{\max}T$; thus, Algorithm 1 is guaranteed to converge. Furthermore, Algorithm 1 requires at maximum  $O(N^2)$ iterations, which is significantly less than $O(2^N)$ required by exhaustive search.

Thus, the remaining task for Algorithm 1 is to solve problem (P3). Note that solving (P3) includes determining $\boldsymbol\pi$ and the waypoints $\{\mathbf{q}_{\pi_k}, 1\leq k\leq K\}$. (P3) is essentially equivalent to the TSP with neighborhoods (TSPN), which is known to be NP-hard \cite{ADumitrscu}. To solve (P3), we propose an efficient method for waypoints design based on TSP method and convex optimization. Specifically, the visiting order $\boldsymbol\pi$ for the SNs in $U_K$ is first determined by simply applying the TSP algorithm over the SNs in $U_K$ while ignoring the coverage (disk) region of each SN. Since the initial and final points of the UAV are fixed, $\boldsymbol\pi$ can be obtained by using a variation of the TSP method ({\em No-Return-Given-Origin-And-End} TSP) \cite{YZeng2}. Various algorithms have been proposed to find high-quality
solutions to TSP efficiently, e.g., with time complexity $O(K^2)$ \cite{CRego}. With the visiting order $\boldsymbol\pi$ determined, the optimal waypoints $\mathbf{q}_{\pi_k}$ can be obtained by solving the following problem,
\begin{eqnarray}
\text{(P4):} \min\limits_{\mathbf{q}_{\pi_k}} && \sum_{k=1}^{K+1}\norm{\mathbf{q}_{\pi_k}-\mathbf{q}_{\pi_{k-1}}}\nonumber\\
\text{s.t.} && \norm{\mathbf{q}_{\pi_k}-\mathbf{w}_{\pi_k}}\leq r_{\pi_k}, 1\leq k\leq K.
\end{eqnarray}

\begin{algorithm}[h]
\caption{Trajectory design algorithm for (P3)}
\begin{algorithmic}[1]
\State\textbf{Input}:$U_K$;
\State Obtain visiting order $\boldsymbol\pi$ by using the {\em No-Return-Given-Origin-And-End} TSP method \cite{YZeng2};
\State Solve (P4) to obtain $\mathbf{q}_{\pi_k}$ and $d_K$;
\State Construct trajectory $\mathbf{Q}$ based on $\boldsymbol\pi$ and $\mathbf{q}_{\pi_k}$ with line segments;
\State {$\mathbf{Q}_0\leftarrow \mathbf{Q}$}; {$d_0\leftarrow d_K$};
\State\textbf{Output}:$\mathbf{Q}_0$, $d_0$;
\label{code:recentEnd}
\end{algorithmic}
\end{algorithm}\vspace{-0.2in}

Note that the objective function of (P4) is a convex
function with respect to $\mathbf{q}_{\pi_k}$, and the coverage area $D_{\pi_k}$ is a convex set. Thus, (P4) is a convex
optimization problem, which can be solved by
standard convex optimization techniques or existing software
such as CVX \cite{MGrant}, with polynomial complexity. Let $d_K=\sum_{k=1}^{K+1}\norm{\mathbf{q}_{\pi_k}-\mathbf{q}_{\pi_{k-1}}}$. The trajectory design algorithm for (P3) with given $U_K$ is summarized in Algorithm 2.

\section{Numerical Results}\label{Problem}
\begin{figure}[h]
\includegraphics[width=3.2in]{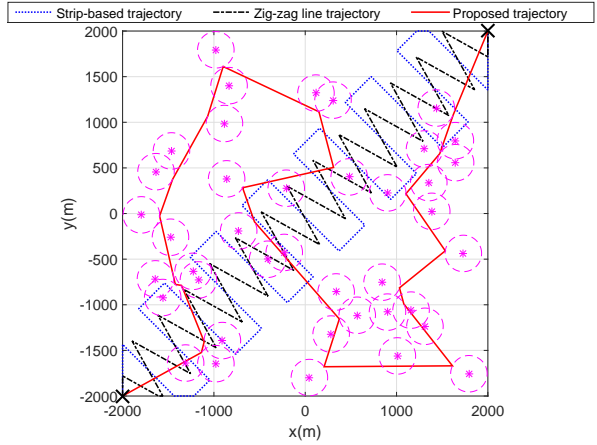}\centering
\caption{Comparison of different UAV trajectories with $T=400$ s.} \label{scheme}\vspace{-0.2in}
\end{figure}

We consider a WSN with $N=40$ SNs, which are randomly located within an area of size $4.0$ km $\times$ $4.0$ km. The following results are based on one random realization of the SN locations as shown in Fig. \ref{scheme}. The UAV's initial and final locations are respectively set as $\mathbf{q}_0=[-2\text{km}, -2\text{km}]^T$ and $\mathbf{q}_{F}=[2\text{km}, 2\text{km}]^T$, and $V_{\max}$ is set as $50$m/s. We assume that the communication range $r_n$ of different SNs is identical, i.e., $r_n=r, \forall n$. If not stated otherwise, we set $r=200$ m. For performance comparison, we also consider two benchmark schemes, namely \textit{strip-based} and \textit{zig-zag line} trajectories for the UAV, as described in the following.

For the strip-based trajectory, the area of interest is partitioned into rectangular strips that are perpendicular to the line connecting $\mathbf{q}_0$ and $\mathbf{q}_F$. Furthermore, each strip has width $2\min_n{r_n}=2r$ so that all SNs within the strip will be visited as the UAV travels along the strip, as shown in Fig. \ref{scheme}. If the rectangular strips exceed the boundary of the area of interest, then the UAV just travels along the intersection between the borderlines of the area and these rectangular strips, which can be uniquely determined. With such strip-based trajectory, the number of visited SNs increases with the height of the strips. Therefore, a bisection search method can be used to determine the maximum height of the strips so that the total UAV flying distance is no greater than $V_{\max}T$. On the other hand, the zig-zag line trajectory is similar to the strip-based trajectory, but with the difference in that rectangular strips are replaced with zig-zag lines. The two benchmarks lead to rather intuitive UAV trajectories for different values of $T$. For example, when $T$ is small, say $T=T_{\min}\triangleq\frac{\norm{\mathbf{q}_F-\mathbf{q}_0}}{V_{\max}}$, the two benchmarks yield the same path that directly connects $\mathbf{q}_0$ and $\mathbf{q}_F$. When $T$ increases, the heights of the strips and zig-zag lines increase since the width of the strip and parallel zig-zag lines are fixed as $2r$. Therefore, when $T$ is sufficiently large, both benchmark schemes result in UAV trajectories covering the entire area of interest, so that all SNs will be visited by the UAV.

\begin{figure}[h]
\subfigure[$K$ v.s. $T$ (with $r=200$ m) ]{\begin{minipage}{1.0\linewidth}\centering
{\includegraphics[width=2.6in]{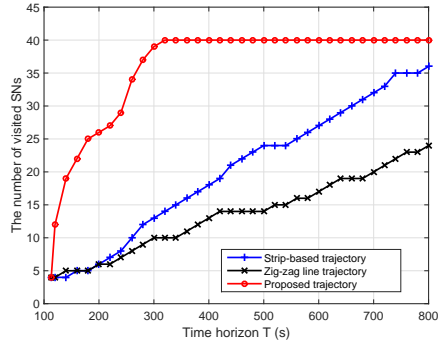}\vspace{0.04in}}
\end{minipage}}
\subfigure[$K$ v.s. $r$ (with $T=200$ s)]{\begin{minipage}{1.0\linewidth}\centering
{\includegraphics[width=2.6in]{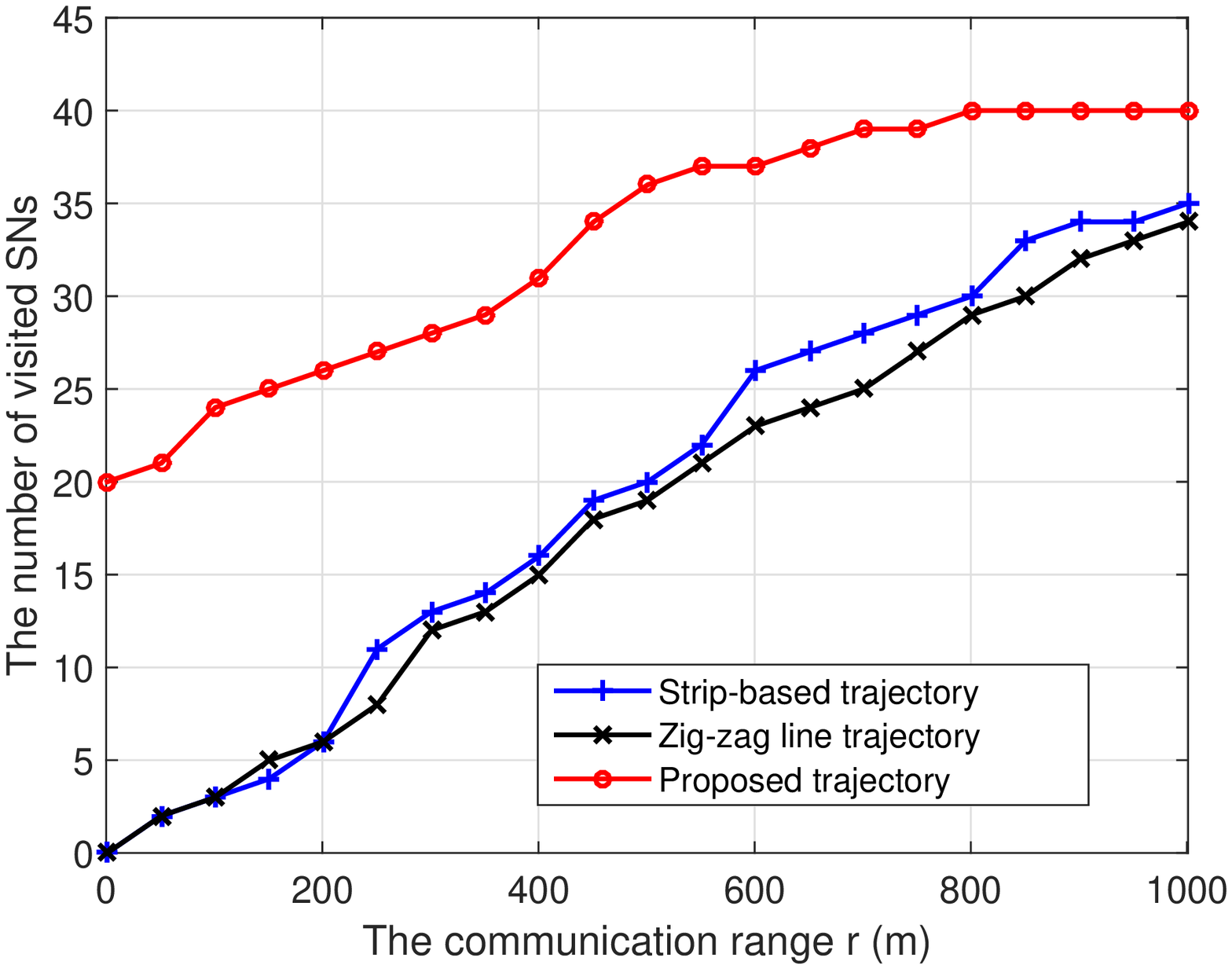}\vspace{0.06in}}
\end{minipage}}\vspace{-0.04in}
\caption{The number of visited SNs $K$ versus the time horizon $T$ or communication range $r$, with $r_n=r, \forall n.$} \label{DifferentTR}
\end{figure}

The optimized
trajectories with different schemes with $T=400$ s are shown in Fig. \ref{scheme}.  It is observed that with our proposed solution, the UAV can visit more SNs than the two benchmark schemes. In Fig. \ref{DifferentTR}(a), we compare the number of visited SNs by
our optimized trajectory with the two benchmark trajectories for different $T$. As expected, our
proposed design significantly outperforms both benchmarks. It is observed that the performance gain is more substantial with small $T>T_{\min}$. As $T$ becomes sufficiently large, all the three trajectories can visit all SNs, but our proposed scheme requires much less time to visit all SNs. It is also observed that the strip-based trajectory gives better performance than the zig-zag line trajectory. This is because the zig-zag line trajectory in general has smaller coverage areas than the strip-based trajectory with the same traveling distance or $T$ (see Fig. \ref{scheme}).\vspace{-0.2in}

Furthermore, we study the effect of the SNs' communication range $r$ on the system performance. Fig. \ref{DifferentTR}(b) plots the number of visited SNs versus $r$ when $T=200$ s. It is observed that for all the three schemes, the number of visited SNs
increases with $r$, as expected; and the proposed trajectory outperforms the two benchmarks significantly, especially for small $r$.

\section{Conclusion}
This paper studies the trajectory design for distributed estimation in a UAV-enabled WSN to minimize the MSE for the estimation, which is shown equivalent to maximize the number of SNs with successful data collection by the UAV.
Although the formulated problem is NP-hard, we reveal that
the optimal UAV trajectory consists of connected line segments only. We then propose a greedy algorithm with low complexity based on TSP method and convex optimization to obtain a suboptimal trajectory solution. Numerical results demonstrate that the proposed design significantly improves the number of visited SNs and hence the estimation performance, as compared to the benchmark schemes.

\end{document}